\documentclass[12pt]{article}
\pdfoutput = 1
\usepackage{caption}
\usepackage{booktabs}
\usepackage{subcaption}
\usepackage{graphics}
\usepackage{graphicx} 
\usepackage{amsmath}
\usepackage{amsfonts}   
\usepackage{amssymb}
\usepackage{hhline}
\usepackage{float}
\usepackage{hyperref}
\begin{document}
\date{Today}
\title{{\bf{\Large Phase transitions in Born-Infeld AdS black holes in $D$-dimensions }}}
\author{{\bf {\normalsize Neeraj Kumar}$^{a}$\thanks{nkneeraj06@gmail.com, @neeraj.kumar@boson.bose.res.in}},\,
{\bf {\normalsize Sukanta Bhattacharyya}$^{b}$
\thanks{sukanta706@gmail.com}},\,
	{\bf {\normalsize Sunandan Gangopadhyay}$^{a}
		$\thanks{Corresponding Author : sunandan.gangopadhyay@gmail.com, sunandan.gangopadhyay@bose.res.in}}\\
	$^{a}$ {\normalsize Department of Theoretical Sciences, S.N. Bose National Centre for Basic Sciences,}\\{\normalsize JD Block, 
		Sector III, Salt Lake, Kolkata 700106, India}\\[0.1cm]
	$^{b}$ {\normalsize Department of Physics, West Bengal State University, Barasat, West Bengal, India}\\[0.1cm]}
\date{}

\maketitle

\begin{abstract}
\noindent In this paper, we have investigated phase transitions in arbitrary spacetime dimensions for Born-Infeld AdS black holes. The phase transition points are characterised from the divergence of heat capacity of the black hole. Two well established techniques, namely, the Ehrenfest scheme and the Ruppeiner state space geometry approach are used to identify the order of phase transition the black hole undergoes. It is observed that the results obtained from these two methods agree with each other. Our analysis reveals that the phase transition is of second order. It is also observed from the variation of the heat capacity with entropy that the small unstable black hole phase becomes more and more stable with increase in the spacetime dimensions. We speculate that this dependence of the stability of the black hole on the spacetime dimension can put an upper limit to the dimension of spacetime from the physical condition of the improbability of the formation of a small stable black hole.
We have also derived a Smarr relation in $D$-spacetime dimensions using scaling arguments and first law of black hole thermodynamics which includes the cosmological constant and the Born-Infeld parameters as thermodynamic variables.
	
\end{abstract}

\vskip 1cm

\section{Introduction}
Black holes are the most exotic objects in physics that behave like a thermodynamical system \cite{aa}-\cite{ac}. It has also been speculated that this surprising connection between black hole and thermodynamics may serve as a new window to quantum gravity. Therefore, the study of black hole thermodynamics received renewed attention in the last two decades. The phenomenon of phase transition in black hole thermodynamics has intrigued researchers working in the field for quite some time and it has been extensively studied as well.  
This phenomenon was first observed in Schwarzchild AdS \cite{ad} background which shows a phase transition between the Schwarzschild AdS black hole to thermal AdS spacetime. Further, the study of black hole thermodynamics in AdS space has become desirable if not absolutely essential due to the discovery of the AdS/CFT duality \cite{jmm}. The key point in the description of phase transition of a black hole is similar to standard thermodynamics, that is, at the critical point(s) some physical quantity changes abruptly. Hence a natural question arises about the order of the phase transition. To address this issue, the  Ehrenfest scheme which is well trusted in standard thermodynamics,  has been applied in \cite{ae}-\cite{robin} for different types of black holes. The uniqueness of the Ehrenfest scheme to classify the nature of the phase transition and the definition of a new parameter called the Prigogine-Defay (PD) \cite{ks},\cite{jp} ratio for the degree of deviation of the second order phase transition (if the phase transition is not truly  second order), makes the Ehrenfest scheme a wonderful tool in standard thermodynamics. This gives enough motivation to apply this scheme in black hole thermodynamics also.
There is yet another interesting approach to tackle the issue of phase transition in thermodynamics. This is the Ruppeiner's thermodynamics state space approach \cite{aq}- \cite{r1}. The method uses the idea that the Hessian of the entropy function can be regarded as a metric tensor on the state space. The Riemannian geometry of this metric gives insight into the underlying statistical thermodynamics of the system. 

In this work, we shall investigate the phase transition in the thermodynamics of Born-Infeld AdS black hole in $D$-dimensions using both the Ehrenfest approach and the Ruppeiner state space geometry technique. The analysis of Born-Infeld black holes is important in its own right since Born-Infeld electrodynamics is one of the most important non-linear electromagnetic theory free from infinite self energies of charged point particles that arises in the Maxwell theory \cite{BI}. A further motivation for looking at black holes in AdS spacetime comes from the importance of the AdS/CFT correspondence in studying strongly coupled systems in quantum field theories \cite{malda}. The study of phase transitions in these black holes was carried in \cite{robin} for $D=4$. However, our analysis holds in any spacetime dimensions $D\geq4$ and hence are more general in the sense that one can observe the effects of dimension of spacetime on the thermodynamics of these black holes. The result can also be important from the point of view of shedding some light on the upper limit that $D$ can take considering the improbability of the stability of the small black hole phase.

In this investigation, we first find the singularities in heat capacity of the black hole and apply the Ehrenfest scheme to analytically characterise the order of phase transition. The analysis reveals that  the black hole undergoes second order phase transition irrespective of its dimensions. Then we apply the well established Ruppeiner state space geometry approach to study the nature of the phase transition. We obtain that the scalar curvature diverges at points where the specific heat shows singularities in any dimension. It is also reassuring to note that the known results are recovered for Reissner-Nordstr$\ddot{o}$m AdS black hole \cite{af} in the limit ($b\rightarrow \infty $ and $Q\neq 0$) where $b$ is the Born-Infeld parameter and $Q$ is the charge of the black hole of $D=4$.

\noindent The paper is organized as follows. In section 2, we study the thermodynamics of the Born-Infeld  AdS black hole in $D$-dimensional spacetime. In section 3, we apply the Ehrenfest scheme to investigate the nature of the phase transitions in this system. In section 4, the Ruppeiner state space geometry approach is used to study phase transitions of the black hole geometry. We conclude in section 5.

\section{Thermodynamics of $D$-dimensional Born-Infeld AdS black holes}

In this section we proceed to explore the thermodynamic properties and the nature of phase
transitions of Born-Infeld AdS black holes in $D$-dimensional spacetime.

\noindent We start by writing down the Born-Infeld black hole solution in $D$-dimensional AdS spacetime. The metric of this spacetime with a negative cosmological constant $\Lambda=-\dfrac{(D-1)(D-2)}{2l^2}$ and with $G$ being set to unity and $D\geq4$ reads
\begin{eqnarray}
ds^2=-f(r)dt^2+\dfrac{1}{f(r)}dr^2 +r^2d\Omega^2_{D-2}
\label{a}
\end{eqnarray}
\noindent where metric coefficient is given by
\begin{eqnarray}
f(r)=1-\dfrac{m}{r^{D-3}}+r^2+\dfrac{4b^2r^2}{(D-1)(D-2)}\left(1-\sqrt{1+\dfrac{(D-2)(D-3)q^2}{2b^2r^{2D-4}}}\right) \nonumber\\ 
+\dfrac{2(D-2)q^2}{(D-1)r^{2D-6}}{}_2F_1\left[\dfrac{D-3}{2D-4},\dfrac{1}{2},\dfrac{3D-7}{2D-4},-\dfrac{(D-2)(D-3)q^2}{2b^2r^{2D-4}}\right]
\label{b}
\end{eqnarray}
where the AdS radius $l$ has been set to unity and ${}_2F_1$ is the hypergeometric function. 

\noindent The mass and charge of the black hole solution reads
\begin{eqnarray}
M &=& \dfrac{(D-2)\omega}{16\pi}m \nonumber\\ 
Q&=&\sqrt{2(D-2)(D-3)}\dfrac{\omega}{8\pi}q~,~~~~~~\omega=\dfrac{2\pi^{\dfrac{D-1}{2}}}{\Gamma\dfrac{(D-1)}{2}}~.
\label{c}
\end{eqnarray}

\noindent It can be seen that in the absence of the  Born-Infeld parameter that is in the $b \to \infty$ limit one recovers the well known solution for the Reissner-Nordstr$\ddot{o}$m AdS black hole geometry. The black hole mass ($M$) in terms of horizon radius ($r_+$) is obtained from the condition $f(r_+)=0$ and is given by
\begin{eqnarray}
M=\dfrac{(D-2)\omega}{16\pi}r_+^{D-3}+\dfrac{(D-2)\omega}{16\pi}r_+^{D-1}+\dfrac{b^2\omega}{4\pi (D-1)}r_+^{D-1}\left(1-\sqrt{1+\dfrac{16\pi^2 Q^2}{b^2\omega^2 r_+^{2D-4}}}\right) \nonumber\\
+\dfrac{4\pi(D-2)Q^2}{(D-1)(D-3)\omega r_+^{D-3}}{}_2F_1\left[\dfrac{D-3}{2D-4},\dfrac{1}{2},\dfrac{3D-7}{2D-4},-\dfrac{16\pi Q^2}{b^2\omega^2 r_+^{2D-4}}\right].
\label{d}
\end{eqnarray}

\noindent The Hawking temperature of the black hole spacetime given by eq.(\ref{b}), reads
\begin{eqnarray}
T &=&\dfrac{1}{4\pi}\left(\dfrac{df(r)}{dr}\right)_{r_+} \nonumber\\
&=&\dfrac{1}{4\pi}\left[\dfrac{D-3}{r_+}+(D-1)r_++\dfrac{4b^2r_+}{D-2}\left(1-\sqrt{1+\dfrac{16\pi^2 Q^2}{b^2\omega^2 r_+^{2D-4}}}\right)\right]~.
\label{e}
\end{eqnarray}
\noindent We now use the relation 
\begin{eqnarray}
dM=TdS+\Phi dQ
\label{tr}
\end{eqnarray}
which is the first law of black hole thermodynamics. This form is analogous to the first law of thermodynamics
\begin{eqnarray}
dU=TdS-PdV
\end{eqnarray}
with the identification of the pressure $P$ to the negative of the electrostatic potential $\Phi$, the volume to the charge $Q$ and the internal energy $U$ to the mass of the black hole $M$. From this relation, the black hole entropy $S$ can be obtained as
\begin{eqnarray}
S=\int_0^{r_{+}} \frac{1}{T} \left(\frac{\partial M}{\partial r}\right)_Q dr=\dfrac{\omega}{4}r_+^{D-2}
\label{f}
\end{eqnarray}

\noindent Now substituting $r_{+}$ in terms of the black hole entropy $S$ from the above equation and plugging it in eq.(\ref{e}), the Hawking temperature of the black hole takes the form
\begin{eqnarray}
T=\dfrac{1}{4\pi}\left[(D-3)\left(\dfrac{\omega}{4S}\right)^{\dfrac{1}{D-2}}+(D-1)\left(\dfrac{4S}{\omega}\right)^{\dfrac{1}{D-2}} \right.\nonumber\\ 
 \left.+\dfrac{4b^2}{(D-2)}\left(\dfrac{4S}{\omega}\right)^{\dfrac{1}{D-2}}\left(1-\sqrt{1+\dfrac{\pi^2Q^2}{b^2S^2}}\right)\right]~. 
\label{g}
\end{eqnarray}
With these results in place, we are now able to study the order of phase transition in this black hole. To do this we plot the Hawking temperature ($T$) with the entropy ($S$) in different dimensions. From the plots, we see that the $T-S$ relation is continuous. This indicates that there is no first order phase transition. However, in all the graphs there is a change in the sign of the slope which indicates the higher order derivative of $T$ being zero. Hence we now proceed to investigate the possibility of the higher order phase transitions in this black hole spacetime.  

\begin{figure}[H]
	\begin{subfigure}{.5 \textwidth}
		\centering
		\includegraphics[width=2.6in]{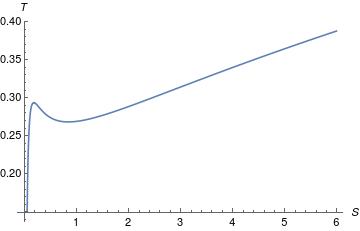}
		\caption{D=4}{(Q=0.13, b=10)}
	\end{subfigure}%
	\begin{subfigure}{.5\textwidth}
  		\centering
  		\includegraphics[width=2.6in]{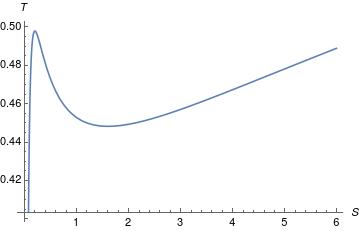}
  		\caption{D=5}{(Q=0.13, b=10)}
	\end{subfigure}
	\begin{subfigure}{.5\textwidth}
  		\centering
  		\includegraphics[width=2.6in]{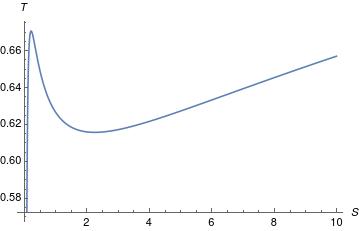}
  		\caption{D=6}{(Q=0.13, b=10)}
	\end{subfigure}
	\caption{Temperature vs Entropy}
\end{figure}
\noindent To investigate this we now calculate the specific heat ($C_{\Phi}$) at constant potential. The gauge potential in $D$ dimensions in case of Born-Infeld black hole geometry is given by
\begin{eqnarray}
\Phi&=& \dfrac{q}{c}\dfrac{1}{r_+^{D-3}}\,
{}_2F_1\left[\frac{D-3}{2 D-4},\frac{1}{2},\frac{3 D-7}{2 D-4},-\frac{(D-2) (D-3)
q^2}{2 b^2r_+^{2D-4}}\right] \\
c&=&\sqrt{\dfrac{2(D-3)}{D-2}}~\nonumber.
\label{h}
\end{eqnarray}
\noindent In terms of $Q$ and $S$, this reads
\begin{eqnarray}
\Phi=\dfrac{4\pi Q}{\omega (D-3)}\left(\dfrac{\omega}{4S}\right)^{\dfrac{D-3}{D-2}}\,{}_2F_1\left[\frac{D-3}{2 D-4},\frac{1}{2},\frac{3 D-7}{2 D-4},-\dfrac{\pi^2 Q^2}{b^2S^2}\right]~.
\label{i}
\end{eqnarray}
\noindent This is the electrostatic potential difference between the event horizon of the black hole and infinity \cite{al}-\cite{ym}. The negative of this potential difference is the analog of the pressure in usual thermodynamics. Considering the temperature of the black hole to be a function of the entropy and charge  ($T\equiv T(S,Q)$), we find 
\begin{eqnarray}
\left(\dfrac{\partial T}{\partial S}\right)_{\Phi}=\left(\dfrac{\partial T}{\partial S}\right)_{Q}-\left(\dfrac{\partial T}{\partial Q}\right)_{S}\left(\dfrac{\partial \Phi}{\partial S}\right)_{Q}\left(\dfrac{\partial Q}{\partial \Phi}\right)_{S}
\label{j}
\end{eqnarray}
where we have the thermodynamic identity
\begin{eqnarray}
\left(\dfrac{\partial Q}{\partial S}\right)_{\Phi}\left(\dfrac{\partial S}{\partial \Phi}\right)_{Q}\left(\dfrac{\partial \Phi}{\partial Q}\right)_{S}=-1
\label{k}
\end{eqnarray}
to obtain eq.(\ref{j}). The heat capacity $C_{\Phi}=T\left(\dfrac{\partial S}{\partial T}\right)_{\Phi}$ can now be calculated using eq.(\ref{g}), (\ref{i}) and (\ref{j}) to be   
\begin{eqnarray}
C_{\Phi}=\dfrac{\mathcal{N}(Q,b,S,\omega)}{\mathcal{M}(Q,b,S,\omega)}
\label{c111}
\end{eqnarray}
where
\begin{eqnarray}
\mathcal{N}(Q,b,S,\omega)=((D-2)S^3)\left(D-3+\sqrt{1+\dfrac{\pi^2Q^2}{b^2S^2}}{}_2F_1\left[\frac{D-3}{2 D-4},\frac{1}{2},\frac{3 D-7}{2 D-4},-\dfrac{\pi^2 Q^2}{b^2S^2}\right]\right)\nonumber\\ 
\times\left[\left(\dfrac{S}{\omega}\right)^{\dfrac{1}{D-2}}2^{\dfrac{4}{D-2}}\left(D^2-3D+2+4b^2\left(1-\sqrt{1+\dfrac{\pi^2Q^2}{b^2S^2}}\right)\right)+\left(\dfrac{\omega}{S}\right)^{\dfrac{1}{D-2}}(D^2-5D+6)\right]
\label{c11}
\end{eqnarray}
\begin{align}
&\mathcal{M}(Q,b,S,\omega)=S^2\left(\dfrac{S}{\omega}\right)^{\dfrac{1}{D-2}}2^{\dfrac{4}{D-2}}\left(D^3-6D^2+11D-6+4b^2(D-3)\left(1-\sqrt{1+\dfrac{\pi^2Q^2}{b^2S^2}}\right)\right)-S^2\nonumber\\
&\times\left(\dfrac{\omega}{S}\right)^{\dfrac{1}{D-2}}(D^3-8D^2+21D-18)+\left(\dfrac{S}{\omega}\right)^{\dfrac{1}{D-2}}2^{\dfrac{4}{D-2}}4\pi^2Q^2(D-3){}_2F_1\left[\frac{D-3}{2 D-4},\frac{1}{2},\frac{3 D-7}{2 D-4},-\dfrac{\pi^2Q^2}{b^2S^2}\right]\nonumber\\
&+S^2\left(\dfrac{S}{\omega}\right)^{\dfrac{1}{D-2}}2^{\dfrac{4}{D-2}}\left[(D^2-3D+2+4b^2)\sqrt{1+\dfrac{\pi^2Q^2}{b^2S^2}}-4b^2\right]{}_2F_1\left[\frac{D-3}{2 D-4},\frac{1}{2},\frac{3 D-7}{2 D-4},-\dfrac{\pi^2 Q^2}{b^2S^2}\right]\nonumber\\
&-S^2\left(\dfrac{\omega}{S}\right)^{\dfrac{1}{D-2}}\sqrt{1+\dfrac{\pi^2Q^2}{b^2S^2}}(D^2-5D+6) {}_2F_1\left[\frac{D-3}{2 D-4},\frac{1}{2},\frac{3 D-7}{2 D-4},-\dfrac{\pi^2 Q^2}{b^2S^2}\right]~.
\label{c12}
\end{align}
 
\noindent We plot this relation between $C_{\Phi}$ and $S$ for different values of spacetime dimensions namely $D=4,~5~,6$. The qualitative nature of the phase transitions of the Born-Infeld AdS black hole in various dimensions can be explored from these plots. The plots have been displayed in Figure 2.

\noindent These reveal that there are two discontinuities in $C_{\Phi}$ and hence refer to the critical points of the phase transition. 
A careful look at these plots reveal that there are three phases of the black hole, namely Phase I ($0<S<S_1$), phase II ($S_1<S<S_2$) and phase III ($S > S_2$).
We know that the sign of the specific heat indicates the stability of the black hole. From the plots we can identify the three phases as the small unstable black hole (SUB), intermediate unstable black hole (IUB) and large stable black hole (LSB) respectively. Reassuringly, for $D=4$, we recover the results in \cite{robin}.

\begin{figure}[H]
\begin{subfigure}{.5\textwidth}
  \centering
  \includegraphics[width=2.7in]{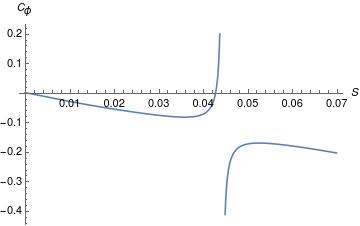}
  \caption{D=4}{(Q=0.13, b=10)}{Critical point($S_{1}$)}
  \label{fig:test11}
\end{subfigure}%
\begin{subfigure}{.5\textwidth}
  \centering
  \includegraphics[width=2.7in]{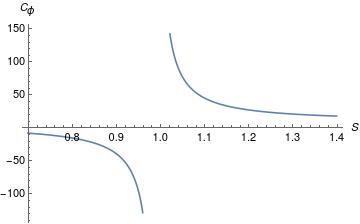}
  \caption{D=4}{(Q=0.13, b=10)}{Critical point ($S_{2}$)}
  \label{fig:test21}
\end{subfigure}
\begin{subfigure}{.5\textwidth}
  \centering
  \includegraphics[width=2.7in]{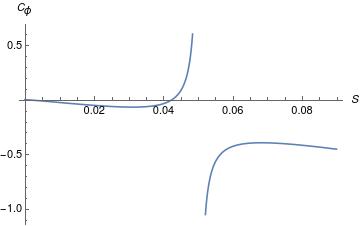}
  \caption{D=5}{(Q=0.13, b=10)}{Critical point ($S_{1}$)}
  \label{fig:test12}
\end{subfigure}%
\begin{subfigure}{.5\textwidth}
  \centering
  \includegraphics[width=2.7in]{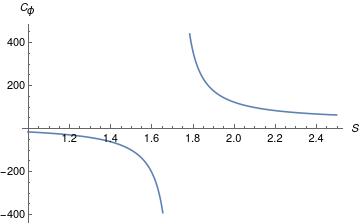}
  \caption{D=5}{(Q=0.13, b=10)}{Critical point ($S_{2}$)}
  \label{fig:test22}
\end{subfigure}
\begin{subfigure}{.5\textwidth}
  \centering
  \includegraphics[width=2.7in]{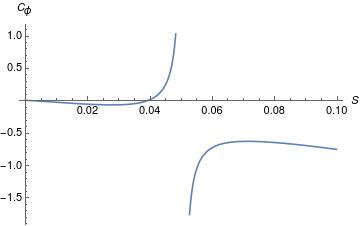}
  \caption{D=6}{(Q=0.13, b=10)}{Critical point ($S_{1}$)}
  \label{fig:test13}
\end{subfigure}%
\begin{subfigure}{.5\textwidth}
  \centering
  \includegraphics[width=2.7in]{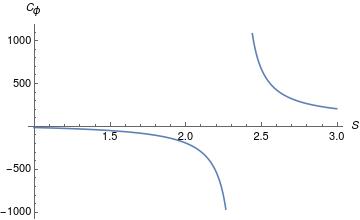}
  \caption{D=6}{(Q=0.13, b=10)}{Critical point ($S_{2}$)}
  \label{fig:test23}
\end{subfigure}
\caption{Heat capacity vs Entropy}
\end{figure}

\noindent  

In Table 1, we have computed the values of entropy where $C_{\Phi}$ diverges for $D=4,5,6$. This we have done by making use of eq.(s)(\ref{c111},\ref{c11},\ref{c12}).

\begin{center}
\begin{tabular}{ |c|c|c|c| }
\hline
 	   & D=4       & D=5       & D=6        \\  	   
\hline 
 $S_1$   & 0.0443255 & 0.0502572 & 0.0506412  \\ 
 $S_2$   & 0.991135  & 1.72042   & 2.35452  \\
 \hline
\end{tabular}
\captionof{table}{Values of entropy where $C_{\Phi}$ diverge.}
\end{center}

\noindent Note that the discontinuities in the specific heat do not necessarily imply second order phase transitions. One can only infer that the order of phase transition is greater than first order. To find out whether the phase transition is of second order, we take recourse to the Ehrenfest scheme and Ruppeiner's curvature method in state space geometry of thermodynamics for the Born-Infeld AdS black hole in $D$-spacetime dimensions. This we do in the subsequent sections.\

Before ending this section, we would like to discuss about the possibility of interpreting the cosmological constant $\Lambda$ as a thermodynamic pressure $p$ and treating the Born-Infeld parameter $b$ as a new thermodynamic variable, as it has been proposed recently in \cite{dolan}, \cite{mann}. The argument put forward in these studies was that since $\Lambda$ and $b$ are dimensionful quantities, the corresponding terms would definitely appear in the Smarr formula. To write down the Smarr formula, we first note that the first law of black hole thermodynamics takes the form \cite{mann}
\begin{eqnarray}
dM=TdS+vdp+\Phi dQ+\mathcal{B}db
\label{fl}
\end{eqnarray}
where $\mathcal{B}$ is the quantity conjugate to $b$ defined by
\begin{eqnarray}
\mathcal{B}=\dfrac{\partial M}{\partial b}~.
\end{eqnarray}
Considering the black hole mass $M=M(S,Q,p,b)$ and performing dimensional analysis, we find that $[M]=L^{D-3},~~[S]=L^{D-2},~~[p]=L^{-2},~~[Q]=L^{D-3}$ and $[b]=L^{-1}$. Using these along with Euler's theorem\footnote{Given a function $g(x,y)$ satisfying $g(\alpha^m x,\alpha^n y)=\alpha^rg(x,y)$, we have $rg(x,y)=mx\left(\dfrac{\partial g}{\partial x}\right)+ny\left(\dfrac{\partial g}{\partial y}\right)$.}, we obtain
\begin{eqnarray}
(D-3)M=(D-2)S\left(\dfrac{\partial M}{\partial S}\right)-2p\left(\dfrac{\partial M}{\partial p}\right)+(D-3)Q\left(\dfrac{\partial M}{\partial Q}\right)-b\left(\dfrac{\partial M}{\partial b}\right)~.
\label{m1}
\end{eqnarray}
Now using eq.(\ref{fl}), we get
\begin{eqnarray}
\left(\dfrac{\partial M}{\partial S}\right)=T,~~\left(\dfrac{\partial M}{\partial p}\right)=v,~~
\left(\dfrac{\partial M}{\partial Q}\right)=\Phi,~~
\left(\dfrac{\partial M}{\partial b}\right)=\mathcal{B}~~.
\end{eqnarray}
Substituting this in eq.(\ref{m1}) yields the Smarr formula
\begin{eqnarray}
M=\dfrac{D-2}{D-3}TS-\dfrac{2}{D-3}pv+\Phi Q-\dfrac{1}{D-3}b\mathcal{B}~~.
\label{sm}
\end{eqnarray}
Reassuringly, the above result reduces to the corresponding relation in $D=4$ spacetime dimensions \cite{mann}
\begin{eqnarray}
M=2TS-2pv+\Phi Q-b\mathcal{B}~~.
\end{eqnarray} 
One can easily verify eq.{\ref{sm}} by computing the left and right hand sides of the relation. 

It would be interesting to work with the first law of black hole thermodynamics given in eq.(\ref{fl}) and study phase transitions. However, we shall not carry out this investigation in this work and we intend to do it in future.

\section{Analysis of phase transition using Ehrenfest scheme}
Ehrenfest's formalism of studying phase transitions is standard technique in thermodynamics to determine the nature of phase transitions for various thermodynamical systems \cite{tm}, \cite{tm2}. It simply says that the order of the phase transition corresponds to the discontinuity in the order of the derivative of Gibb's potential. The Maxwell's relations in thermodynamics give the conditions to be satisfied for a particular phase transition. \\

\noindent The first and second Ehrenfest equations in thermodynamics are given by
\begin{eqnarray}
\left(\dfrac{\partial P}{\partial T}\right)_S&=&\dfrac{1}{VT}\dfrac{C_{P_2}-C_{P_1}}{\beta_2-\beta_1}=\dfrac{\Delta C_P}{VT\Delta\beta},
\label{eq1}
\end{eqnarray}
\begin{eqnarray}
\left(\dfrac{\partial P}{\partial T}\right)_V&=&\dfrac{\beta_2-\beta_1}{\kappa_2-\kappa_1}=\dfrac{\Delta\beta}{\Delta\kappa}
\label{eq2}
\end{eqnarray}
\noindent where subscripts $1$ and $2$ denote two distinct phases of the system. Now we use the correspondence between the pressure ($P$) to the negative of the electrostatic potential difference ($\Phi$) and the volume ($V$) to the charge ($Q$) of the black hole. These identifications lead to the following equations 
\begin{eqnarray}
-\left(\dfrac{\partial \Phi}{\partial T}\right)_S&=&\dfrac{1}{QT}\dfrac{C_{{\Phi}_2}-C_{\Phi_1}}{\beta_2-\beta_1}=\dfrac{\Delta C_{\Phi}}{QT\Delta\beta}
\label{ee1}
\end{eqnarray}
\begin{eqnarray}
-\left(\dfrac{\partial \Phi}{\partial T}\right)_Q&=&\dfrac{\beta_2-\beta_1}{\kappa_2-\kappa_1}=\dfrac{\Delta\beta}{\Delta\kappa}~.
\label{ee2}
\end{eqnarray}
Note that $\beta$ is the volume expansion coefficient and $\kappa$ is the isothermal compressibility of the system and are defined as
\begin{eqnarray}
\beta=\dfrac{1}{Q}\left(\dfrac{\partial Q}{\partial T}\right)_{\Phi}~,~~~~
\kappa=\dfrac{1}{Q}\left(\dfrac{\partial Q}{\partial \Phi}\right)_{T}~.
\label{veic}
\end{eqnarray}
\noindent Now we proceed to check whether the black hole phase transition satisfies the Ehrenfest equations (\ref{ee1}, ref{ee2}). In other words, we investigate the validity of the Ehrenfest equations at the points of discontinuities $S_i$,~(i=1, 2). Here we denote the critical values of temperature by $T_i$ and charge by $Q_i$.

\noindent The left hand side of the first Ehrenfest equation (\ref{ee2}) at the critical point can be written as
\begin{eqnarray}
-\left[\left(\dfrac{\partial \Phi}{\partial T}\right)_S\right]_{S=S_i}&=&-\left[\left(\dfrac{\partial \Phi}{\partial Q}\right)_S\right]_{S=S_i}\left[\left(\dfrac{\partial Q}{\partial T}\right)_S\right]_{S=S_i} \nonumber\\ 
&=&-\dfrac{\left[\left(\dfrac{\partial \Phi}{\partial Q}\right)_S\right]_{S=S_i}}{\left[\left(\dfrac{\partial T}{\partial Q}\right)_S\right]_{S=S_i}}~.
\label{t}
\end{eqnarray}

\noindent Using eq.(\ref{veic}) and the definition of heat capacity $C_{\Phi}= T(\frac{\partial S}{\partial T})_{\Phi}$, we can obtain the right hand side of eq.(\ref{ee1}) to be 
\begin{equation} 
Q_i\beta=\left[\left(\dfrac{\partial Q}{\partial T}\right)_{\Phi}\right]_{S=S_i}=\left[\left(\dfrac{\partial Q}{\partial S}\right)_{\Phi}\right]_{S=S_i}\left(\dfrac{C_{\Phi}}{T_i}\right)
\end{equation}

\begin{equation}
\dfrac{\Delta C_{\Phi}}{T_i Q_i\Delta\beta}=\left[\left(\dfrac{\partial S}{\partial Q}\right)_{\Phi}\right]_{S=S_i}~.
\end{equation}
\noindent Using the identity (\ref{k}), the above equation can be written in the form

\begin{eqnarray}
\dfrac{\Delta C_{\Phi}}{T_i Q_i\Delta\beta}=-\dfrac{
\left[\left(\dfrac{\partial \Phi}{\partial Q}\right)_S\right]_{S=S_i}}{\left[\left(\dfrac{\partial \Phi}{\partial S}\right)_Q\right]_{S=S_i}}~.
\label{u}
\end{eqnarray}
Eq.(s) (\ref{ee1}, \ref{t}, \ref{u}) give the first Ehrenfest's equation to be
\begin{eqnarray}
\left[\left(\dfrac{\partial T}{\partial Q}\right)_S\right]_{S=S_i}=\left[\left(\dfrac{\partial \Phi}{\partial S}\right)_Q\right]_{S=S_i}~.
\label{40}
\end{eqnarray}
This is another form of the first Ehrenfest equation given in eq.(\ref{ee1}). In the subsequent discussion we shall show the validity of the above form of the Ehrenfest equation.
Differentiating eq.(s)(\ref{g}, \ref{h}), we get the left hand side of the above equation to be  
\begin{eqnarray}
\left[\left(\dfrac{\partial T}{\partial Q}\right)_S\right]_{S=S_i}=-\dfrac{\pi Q_i}{(D-2)S_i^2}\left(\dfrac{4S_i}{\omega}\right)^{\dfrac{1}{D-2}}\left(1+\dfrac{\pi^2Q_i^2}{b^2S_i^2}\right)^{-\dfrac{1}{2}}~
\label{31}
\end{eqnarray}
and right hand side of eq.(\ref{40}) to be
\begin{eqnarray}
\left[\left(\dfrac{\partial \Phi}{\partial S}\right)_Q\right]_{S=S_i}=-\dfrac{\pi Q_i}{(D-2)S_i^2}\left(\dfrac{4S_i}{\omega}\right)^{\dfrac{1}{D-2}}\left(1+\dfrac{\pi^2Q_i^2}{b^2S_i^2}\right)^{-\dfrac{1}{2}}~.
\label{33}
\end{eqnarray}
This reveals that the black hole spacetime follows the first Ehrenfest's equation.\\
One can also check the validity of the first Ehrenfest equation by calculating $\left[\left(\dfrac{\partial \Phi}{\partial Q}\right)_S\right]_{S=S_i}$ which reads
\begin{eqnarray}
\left[\left(\dfrac{\partial \Phi}{\partial Q}\right)_S\right]_{S=S_i}=\dfrac{4\pi}{\omega (D-2)(D-3)}\left(\dfrac{\omega}{4S_i}\right)^{\dfrac{D-3}{D-2}}\left[(D-3)\left(1+\dfrac{\pi^2Q_i^2}{b^2S_i^2}\right)^{-\dfrac{1}{2}} \right.\nonumber\\ 
 \left.+\,{}_2F_1\left[\frac{D-3}{2 D-4},\frac{1}{2},\frac{3 D-7}{2 D-4},-\dfrac{\pi^2 Q_i^2}{b^2S_i^2}\right]\right]~.
 \label{32}
\end{eqnarray}
Therefore eq.(\ref{t}) when calculated using eq.(s)(\ref{31}, \ref{32}) gives
\begin{eqnarray}
-\left[\left(\dfrac{\partial \Phi}{\partial T}\right)_S\right]_{S=S_i}=\dfrac{S_i}{Q_i}\left[1+\dfrac{1}{D-3}\left(1+\dfrac{\pi^2Q_i^2}{b^2S_i^2}\right)^{\dfrac{1}{2}}\,{}_2F_1\left[\frac{D-3}{2 D-4},\frac{1}{2},\frac{3 D-7}{2 D-4},-\dfrac{\pi^2 Q_i^2}{b^2S_i^2}\right]\right]~.
\label{101}
\end{eqnarray}
Now eq.(\ref{u}) when calculated using eq.(s)(\ref{33}, \ref{32}) gives
\begin{eqnarray}
\dfrac{\Delta C_{\Phi}}{T_i Q_i\Delta\beta}=\dfrac{S_i}{Q_i}\left[1+\dfrac{1}{D-3}\left(1+\dfrac{\pi^2Q_i^2}{b^2S_i^2}\right)^{\dfrac{1}{2}}\,{}_2F_1\left[\frac{D-3}{2 D-4},\frac{1}{2},\frac{3 D-7}{2 D-4},-\dfrac{\pi^2 Q_i^2}{b^2S_i^2}\right]\right]~.
\label{102}
\end{eqnarray}
Eq.(s)(\ref{101}, \ref{102}) once again show the validity of the first Ehrenfest equation (\ref{ee1}). It is reassuring to note that eq.(s)(\ref{101}, \ref{102}) reduce to the expression in \cite{robin} for $D=4$
\begin{eqnarray}
-\left[\left(\dfrac{\partial \Phi}{\partial T}\right)_S\right]_{S=S_i}=\dfrac{\Delta C_{\Phi}}{T_i Q_i\Delta\beta}=\dfrac{S_i}{Q_i}\left[1+\left(1+\dfrac{\pi^2Q_i^2}{b^2S_i^2}\right)^{\dfrac{1}{2}}\,{}_2F_1\left[\dfrac{1}{4},\dfrac{1}{2},\dfrac{5}{4},-\dfrac{\pi^2 Q_i^2}{b^2S_i^2}\right]\right]~.
\end{eqnarray}
Now we focus our attention to the second Ehrenfest equation (\ref{ee2}). To calculate the left hand side of the second Ehrenfest equation, we use the thermodynamic relation $$T\equiv T(S,\Phi)$$ 
\begin{eqnarray}
\left(\dfrac{\partial T}{\partial \Phi}\right)_Q=\left(\dfrac{\partial T}{\partial S}\right)_{\Phi}\left(\dfrac{\partial S}{\partial \Phi}\right)_Q+\left(\dfrac{\partial T}{\partial \Phi}\right)_S~.
\end{eqnarray}
Since heat capacity diverges at the critical points, hence $\left[\left(\dfrac{\partial T}{\partial S}\right)_{\Phi}\right]_{S=S_i}=0$, and $\left(\dfrac{\partial S}{\partial \Phi}\right)_Q$ is finite at the critical point which can be inferred from eq.(\ref{33}), therefore,
\begin{eqnarray}
\left[\left(\dfrac{\partial T}{\partial \Phi}\right)_Q\right]_{S=S_i}=\left[\left(\dfrac{\partial T}{\partial \Phi}\right)_S\right]_{S=S_i}~.
\end{eqnarray}
This in turn implies
\begin{eqnarray}
\left[\left(\dfrac{\partial \Phi}{\partial T}\right)_Q\right]_{S=S_i}=\left[\left(\dfrac{\partial \Phi}{\partial T}\right)_S\right]_{S=S_i}=-\dfrac{S_i}{Q_i}\left[1+\dfrac{1}{D-3}\left(1+\dfrac{\pi^2Q_i^2}{b^2S_i^2}\right)^{\dfrac{1}{2}}\,{}_2F_1\left[\frac{D-3}{2 D-4},\frac{1}{2},\frac{3 D-7}{2 D-4},-\dfrac{\pi^2 Q_i^2}{b^2S_i^2}\right]\right]
\label{103}
\end{eqnarray}
where we have used eq.(\ref{101}) in the second equality of the above equation.\

\noindent Now from eq.(\ref{veic}), at the critical points
\begin{eqnarray}
\kappa Q_i=\left[\left(\dfrac{\partial Q}{\partial \Phi}\right)_T\right]_{S=S_i}~.
\end{eqnarray}
Using the thermodynamic identity $\left(\dfrac{\partial Q}{\partial \phi}\right)_T\left(\dfrac{\partial \Phi}{\partial T}\right)_Q\left(\dfrac{\partial T}{\partial Q}\right)_{\Phi}=-1$ and the definition of $\beta$ in eq.(\ref{veic}), we find
\begin{eqnarray}
\kappa Q_i=\left[\left(\dfrac{\partial T}{\partial \Phi}\right)_Q\right]_{S=S_i}Q_i\beta~.
\end{eqnarray}
\noindent Therefore, the right hand side of the second Ehrenfest equation (\ref{ee2}) reduces to 
\begin{eqnarray}
\dfrac{\Delta \beta}{\Delta \kappa}=-\left[\left(\dfrac{\partial \Phi}{\partial T}\right)_Q\right]_{S=S_i}~.
\end{eqnarray}
This can further be written as
\begin{eqnarray}
-\left[\left(\dfrac{\partial \Phi}{\partial T}\right)_Q\right]_{S=S_i}&=&-\left[\left(\dfrac{\partial \Phi}{\partial S}\right)_Q\right]_{S=S_i}\left[\left(\dfrac{\partial S}{\partial T}\right)_Q\right]_{S=S_i}\nonumber\\
&=&-\dfrac{\left[\left(\dfrac{\partial \Phi}{\partial S}\right)_Q\right]_{S=S_i}}{\left[\left(\dfrac{\partial T}{\partial S}\right)_Q\right]_{S=S_i}}~.
\label{35}
\end{eqnarray}
Now from eq.(\ref{g}) we have
\begin{eqnarray}
\left[\left(\dfrac{\partial T}{\partial S}\right)_Q\right]_{S=S_i}=\dfrac{1}{4\pi}\left[-\dfrac{(D-3)}{(D-2)S_i}\left(\dfrac{\omega}{4S_i}\right)^{\dfrac{1}{D-2}}+\dfrac{(D-1)}{(D-2)S_i}\left(\dfrac{4S_i}{\omega}\right)^{\dfrac{1}{D-2}}\dfrac{4b^2}{(D-2)^2S}\right.\nonumber\\ 
 \times\left.\left(\dfrac{4S_i}{\omega}\right)^{\dfrac{1}{D-2}}\left(1-\sqrt{1+\dfrac{\pi^2Q_i^2}{b^2S_i^2}}\right)+\dfrac{4\pi^2Q_i^2}{(D-2)S^3}\left(\dfrac{4S_i}{\omega}\right)^{\dfrac{1}{D-2}}\left(1+\dfrac{\pi^2Q_i^2}{b^2S_i^2}\right)^{-\dfrac{1}{2}}\right]~.
 \label{34}
\end{eqnarray}
Using eq.(s)(\ref{33}, \ref{34}), eq.(\ref{35}) becomes
\begin{eqnarray}
-\left[\left(\dfrac{\partial \Phi}{\partial T}\right)_Q\right]_{S=S_i}=\dfrac{S_i}{Q_i}\left[1+\dfrac{1}{D-3}\left(1+\dfrac{\pi^2Q_i^2}{b^2S_i^2}\right)^{\dfrac{1}{2}}\,{}_2F_1\left[\frac{D-3}{2 D-4},\frac{1}{2},\frac{3 D-7}{2 D-4},-\dfrac{\pi^2 Q_i^2}{b^2S_i^2}\right]\right]~.
\label{104}
\end{eqnarray}
Eq.(s)(\ref{103},\ref{104}) prove the validity of the second Ehrenfest equation (\ref{ee2}) also at the critical points.\

\noindent To conclude the discussion, we calculate the Prigogine-Defay (PD) ratio. To do so we note that eq.(\ref{103}) gives
\begin{eqnarray}
\left[\left(\dfrac{\partial \Phi}{\partial T}\right)_Q\right]_{S=S_i}=\left[\left(\dfrac{\partial \Phi}{\partial T}\right)_S\right]_{S=S_i}=-\dfrac{\Delta C_{\Phi}}{T_i Q_i\Delta\beta}
\label{105}
\end{eqnarray}
where the first Ehrenfest equation (\ref{ee1}) at the critical point has been used in the second equality. However, by the second Ehrenfest equation (\ref{ee2}) at the critical point
\begin{eqnarray}
\left[\left(\dfrac{\partial \Phi}{\partial T}\right)_Q\right]_{S=S_i}=-\dfrac{\Delta \beta}{\Delta \kappa}~.
\label{106}
\end{eqnarray}
Eq.(s)(\ref{105},\ref{106}) yields the PD ratio to be
\begin{eqnarray}
\Pi=\dfrac{\Delta C_{\Phi}\Delta\kappa}{T_i Q_i(\Delta\beta)^2}=1~.
\end{eqnarray}
The above analysis clearly shows an exact second order nature of the black hole phase transition. In the next section we shall reinforce these results using the Ruppeiner state space geometry approach to thermodynamics.

\section{Phase transition using Ruppeiner state space geometry}
In the previous section, we have seen that the two equations of Ehrenfest are satisfied at the critical points. This suggests that the black hole phase transition is second order in nature. In this section, we analyse the same phenomenon using another technique known as the Ruppeiner thermodynamic state space geometry approach. The  thermodynamics state space geometry approach has been extensively studied to explore the phase transition phenomena in black hole thermodynamics from the past two decades \cite{r3}-\cite{r6}. To proceed further we first calculate the scalar curvature ($R$) of the two dimensional thermodynamic manifold.

The Ruppeiner metric coefficients for the manifold are given by\cite{ar}, \cite{as}
\begin{equation}
g^{R}_{ij}=-\dfrac{\partial^2 S(x^i)}{\partial x^i \partial x^j}
\end{equation}
where $x^i=x^i(M, Q)$; i=1,2 are extensive variables of the manifold.
\noindent The calculation of the Weinhold metric coefficients is convenient for computational purpose. These are given by
\begin{equation}
g^{W}_{ij}=\dfrac{\partial^2 M(x^i)}{\partial x^i \partial x^j}
\end{equation}
where $x^i=x^i(S, Q)$; i=1,2. It is to be noted that Weinhold geometry is connected to the Ruppeiner geometry through the following map
\begin{equation}
dS^2_R=\dfrac{dS^2_W}{T}~.
\end{equation}
The mass of the black hole can be written in terms of the entropy using eq(s).(\ref{d}, \ref{f}) and reads
\begin{eqnarray}
M(S, Q)=\dfrac{(D-2)\omega}{16\pi}\left(\dfrac{4S}{\omega}\right)^{\dfrac{D-3}{D-2}}+\left(\dfrac{4S}{\omega}\right)^{\dfrac{D-1}{D-2}}\left[\dfrac{(D-2)\omega}{16\pi}+\dfrac{b^2\omega}{4\pi(D-1)}\left(1-\sqrt{1+\dfrac{\pi^2 Q^2}{b^2S^2}}\right)\right]\nonumber\\
+\dfrac{4\pi(D-2)}{\omega(D-1)(D-3)}Q^2\left(\dfrac{\omega}{4S}\right)^{\dfrac{D-3}{D-2}}\,_2F_1\left[\frac{D-3}{2 D-4},\frac{1}{2},\frac{3 D-7}{2 D-4},-\dfrac{\pi^2 Q^2}{b^2S^2}\right]~.
\end{eqnarray}
Differentiating the above equation twice, we get the metric coefficients as
\begin{eqnarray}
g^R_{SS}=\dfrac{1}{T}\left[-\dfrac{(D-3)\omega}{16\pi(D-2)}\left(\dfrac{4}{\omega}\right)^{\dfrac{D-3}{D-2}}S^{-\dfrac{D-1}{D-2}}+\dfrac{(D-1)\omega}{16\pi(D-2)}\left(\dfrac{4}{\omega}\right)^{\dfrac{D-1}{D-2}}S^{-\dfrac{D-3}{D-2}}\right.\nonumber\\
\left. +\dfrac{b^2\omega}{4\pi(D-2)^2}\left(\dfrac{4}{\omega}\right)^{\dfrac{D-1}{D-2}}S^{-\dfrac{D-3}{D-2}}\left(1-\sqrt{1+\dfrac{\pi^2 Q^2}{b^2 S^2}}\right)\right. \nonumber\\
\left.+\dfrac{\omega \pi Q^2}{4(D-2)S^3}\left(\dfrac{4}{\omega}\right)^{\dfrac{D-1}{D-2}}S^{\dfrac{1}{D-2}}\left(1+\dfrac{\pi^2 Q^2}{b^2 S^2}\right)^{-\dfrac{1}{2}}\right]
\end{eqnarray}
\begin{eqnarray}
g^R_{SQ}=\dfrac{1}{T}\left[-\dfrac{\omega \pi Q}{4(D-2)S^2}\left(\dfrac{4}{\omega}\right)^{\dfrac{D-1}{D-2}}S^{\dfrac{1}{D-2}}\left(1+\dfrac{\pi^2 Q^2}{b^2 S^2}\right)^{-\dfrac{1}{2}}\right]
\end{eqnarray}
\begin{eqnarray}
g^R_{QQ}=\dfrac{1}{T}\left(\dfrac{4\pi}{\omega (D-3)}\left(\dfrac{\omega}{4S}\right)^{\dfrac{D-3}{D-2}}\left[\dfrac{D-3}{D-2}\left(1+\dfrac{\pi^2 Q^2}{b^2 S^2}\right)^{-\dfrac{1}{2}}\right.\right.\nonumber \\
\left.\left.+\dfrac{1}{D-2}\,_2F_1\left[\frac{D-3}{2 D-4},\frac{1}{2},\frac{3 D-7}{2 D-4},-\dfrac{\pi^2 Q^2}{b^2S^2}\right]\right]\right)
\label{gqq}
\end{eqnarray}
\noindent where $T$ is the Hawking temperature of the black hole and is defined in eq.(\ref{g}).\

 The Ruppeiner's curvature can be calculated with the metric coefficients using the following relation
\begin{eqnarray}
R &=&-\dfrac{1}{\sqrt{g}}\left[\dfrac{\partial}{\partial S}\left(\dfrac{g_{SQ}}{\sqrt{g}g_{SS}}\dfrac{\partial g_{SS}}{\partial Q}-\dfrac{1}{\sqrt{g}}\dfrac{g_{QQ}}{\partial S}\right)+\dfrac{\partial}{\partial Q}\left(\dfrac{2}{\sqrt{g}}\dfrac{\partial  g_{SQ}}{\partial Q}-\dfrac{1}{\sqrt{g}}\dfrac{\partial g_{SS}}{\partial Q}-\dfrac{g_{SQ}}{\sqrt{g}g_{SS}}\dfrac{\partial g_{SS}}{\partial S}\right)\right]\nonumber\\
&=&\dfrac{A(Q, S)}{B(Q, S)}
\end{eqnarray}
where
\begin{eqnarray}
A(Q, S)=-\left[\dfrac{\partial}{\partial S}\left(\dfrac{g_{SQ}}{\sqrt{g}g_{SS}}\dfrac{\partial g_{SS}}{\partial Q}-\dfrac{1}{\sqrt{g}}\dfrac{g_{QQ}}{\partial S}\right)+\dfrac{\partial}{\partial Q}\left(\dfrac{2}{\sqrt{g}}\dfrac{\partial  g_{SQ}}{\partial Q}-\dfrac{1}{\sqrt{g}}\dfrac{\partial g_{SS}}{\partial Q}-\dfrac{g_{SQ}}{\sqrt{g}g_{SS}}\dfrac{\partial g_{SS}}{\partial S}\right)\right]\nonumber
\end{eqnarray}
and $B(Q,S)=\sqrt{g}$.\

The expression for $R$ is too large to be presented, however, the denominator $B(Q,S)$ takes the form
\begin{eqnarray}
B(Q,S)=\dfrac{1}{T}\left[\dfrac{\pi^2Q^2}{(D-2)^2S^3}\left(\dfrac{4}{\omega}\right)^{\dfrac{2}{D-2}}S^{-\dfrac{D-4}{D-2}}\left[\left(1+\dfrac{\pi^2Q^2}{b^2S^2}\right)^{-1}+\dfrac{1}{D-3}\left(1+\dfrac{\pi^2Q^2}{b^2S^2}\right)^{-\dfrac{1}{2}}\textit{F}\right]\right.\nonumber\\ 
 \left.+\dfrac{b^2}{(D-2)^3}\left(\dfrac{4}{\omega}\right)^{\dfrac{2}{D-2}}S^{-\dfrac{2(D-3)}{D-2}}\left(1-\sqrt{1+\dfrac{\pi^2Q^2}{b^2S^2}}\right)\left[\left(1+\dfrac{\pi^2Q^2}{b^2S^2}\right)^{-\dfrac{1}{2}}+\dfrac{1}{D-3}\textit{F}\right] \right.\nonumber\\ 
 \left.
+\dfrac{D-1}{4(D-2)^2}\left(\dfrac{4}{\omega}\right)^{\dfrac{2}{D-2}}S^{-\dfrac{2(D-3)}{D-2}}\left[\left(1+\dfrac{\pi^2Q^2}{b^2S^2}\right)^{-\dfrac{1}{2}}+\dfrac{1}{D-3}\textit{F}\right] \right.\nonumber\\ 
 \left.
-\dfrac{1}{4(D-2)^2S^2}\left[(D-3)\left(1+\dfrac{\pi^2Q^2}{b^2S^2}\right)^{-\dfrac{1}{2}}+\textit{F}\right] \right.\nonumber\\ 
 \left.
-\dfrac{\omega^2\pi^2Q^2}{16(D-2)^2S^4}\left(\dfrac{4}{\omega}\right)^{\dfrac{2(D-1)}{D-2}}S^{\dfrac{2}{D-2}}\left(1+\dfrac{\pi^2Q^2}{b^2S^2}\right)^{-1}\right]^{\dfrac{1}{2}}
\end{eqnarray}
where $\textit{F}=\,_2F_1\left[\frac{D-3}{2 D-4},\frac{1}{2},\frac{3 D-7}{2 D-4},-\dfrac{\pi^2 Q^2}{b^2S^2}\right]$.\

For $D=4$, the above expression reduces to 
\begin{eqnarray}
B(Q,S)\equiv\sqrt{g}=\dfrac{1}{T}\left[\left(\textit{F}+\left(1+\dfrac{\pi^2Q^2}{b^2S^2}\right)^{-\dfrac{1}{2}}\right)\left(\dfrac{3S-\pi}{16\pi S^2}+\dfrac{b^2}{8\pi S}\left(1-\left(1+\dfrac{\pi^2Q^2}{b^2S^2}\right)^{\dfrac{1}{2}}\right)\right.\right.\nonumber\\ 
 \left.\left.
 +\dfrac{\pi Q^2}{4S^3}\left(1+\dfrac{\pi^2Q^2}{b^2S^2}\right)^{-\dfrac{1}{2}}\right)-\dfrac{\pi Q^2}{4S^3}\left(1+\dfrac{\pi^2Q^2}{b^2S^2}\right)^{-1}\right]^{\dfrac{1}{2}}~.
\end{eqnarray}
The zeroes of this expression gives the critical points.
We plot this expression for different dimensions $D=4,5,6$ for values Q=0.13 and b=10.
\begin{figure}[H]
	\begin{subfigure}{.5\textwidth}
	\centering
		\includegraphics[width=2.5in]{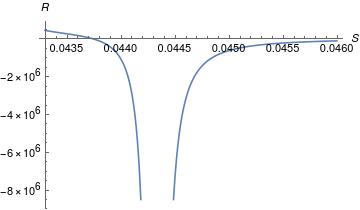}
		\caption{D=4}{(Q=0.13, b=10)}{Critical point ( $S_{1}$)}
		\label{fig:test14}
	\end{subfigure}%
	\begin{subfigure}{.5\textwidth}
	\centering
		\includegraphics[width=2.5in]{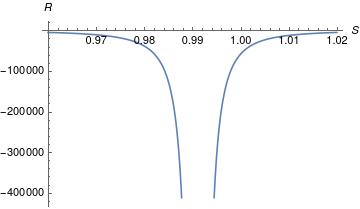}
		\caption{D=4}{(Q=0.13, b=10)}{Critical point ($S_{2}$)}
		\label{fig:test24}
	\end{subfigure}
	\begin{subfigure}{.5\textwidth}
		\centering
		\includegraphics[width=2.5in]{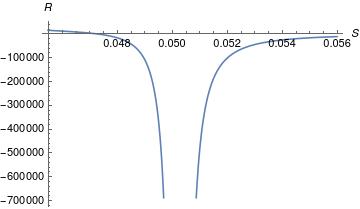}
		\caption{D=5}{(Q=0.13, b=10)}{Critical point ( $S_{1}$)}
		\label{fig:test15}
	\end{subfigure}%
	\begin{subfigure}{.5\textwidth}
		\centering
		\includegraphics[width=2.5in]{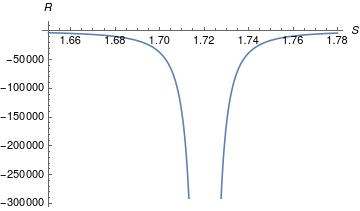}
		\caption{D=5}{(Q=0.13, b=10)}{Critical point ($S_{2}$)}
		\label{fig:test25}
	\end{subfigure}
	\begin{subfigure}{.5\textwidth}
		\centering
		\includegraphics[width=2.5in]{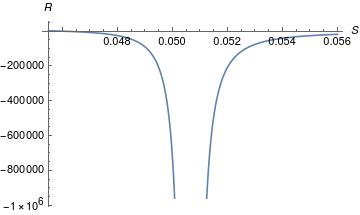}
		\caption{D=6}{(Q=0.13, b=10)}{Critical point ($S_{1}$)}
		\label{fig:test16}
	\end{subfigure}%
	\begin{subfigure}{.5\textwidth}
		\centering
		\includegraphics[width=2.5in]{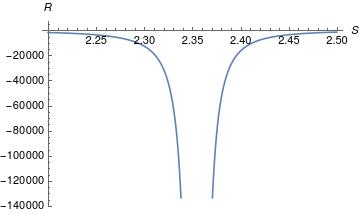}
		\caption{D=6}{(Q=0.13, b=10)}{Critical point ($S_{2}$)}
		\label{fig:test26}
	\end{subfigure}
	\caption{Ruppeiner curvature vs Entropy}
\end{figure}

\noindent From the plots, it can be observed that the Ruppeiner curvature diverges for the same values of entropy as in heat capacity. 


The values of entropy where Ruppeiner curvature diverges are presented in Table 2
\begin{center}
\begin{tabular}{ |c|c|c|c| }
\hline
       & D=4   & D=5  & D=6 \\ 
 \hline 

 $S_1$   & 0.0443255 & 0.0502572 & 0.0506412  \\ 
 $S_2$   & 0.991135  & 1.72042   & 2.35452   \\
 	\hline
\end{tabular}
\captionof{table}{Values of entropy where $R$ diverge.}
\end{center}
Simply comparing the critical points from Table 1 and Table 2, we see that the Ruppeiner cuvature and the heat capacity diverge at the same points $S_1$ and $S_2$. 
\section{Conclusion}
In this paper, we have investigated the phase transition in $AdS$ Born-Infeld black holes in $D$-spacetime dimensions. 
The importance of studying these black holes lies in the fact that Born-Infeld electrodynamics is one of the most important non-linear electromagnetic theory free from infinite self energies of charged point particles. 
We have first obtained expressions for the Hawking temperature, entropy and heat capacity of these black holes in arbitrary spacetime dimensions. We have then presented the dependence of the Hawking temperature with entropy for $D=4,5,6$. The dependence look similar for all the three cases. However, we observe that the peak in the Hawking temperature increases with increase in the spacetime dimensions $D$. We have then plotted the variation of the heat capacity with the entropy for $D=4, 5, 6$. We observe that there are three phases possible in all three cases. These are the small unstable black hole phase, intermediate unstable black hole phase and the large stable black hole phase. It is  seen that the small unstable black hole phase becomes more and more stable as we increase the spacetime dimensions. This could in principle mean that there is an upper limit to the number of spacetime dimensions possible due to the improbability of the formation of the small stabe black hole. We have then used the Ehrenfest scheme and the Ruppeiner state space geometry approach to identify the order of the phase transition. It is found that even for higher dimensions, Ehrenfest scheme as well as the state space geometry approach for this kind of black holes show second order phase transition. To substantiate our observations, we have provided analytical expressions fo relevant thermodynamical quantities around the critical point. The critical points obtained from state space geometry technique agree with the points where the heat capacity diverges. 
Finally, we have also derived a Smarr relation in $D$-spacetime dimensions using scaling arguments and first law of black hole thermodynamics in which the cosmological constant and the Born-Infeld parameters have been treated as thermodynamic variables.

\section*{Acknowledgement}S.G. acknowledges the support of the Visiting Associateship
programme of Inter University Centre for Astronomy and Astrophysics (IUCAA), Pune.
The authors would like to thank the referee for useful comments.

\end{document}